\documentstyle[psfig]{cupconf}

\ifoldfss
\else
  \ifnfssone
    \newmathalphabet{\mathit}
      \addtoversion{normal}{\mathit}{cmr}{m}{it}
      \addtoversion{bold}{\mathit}{cmr}{bx}{it}
    \newmathalphabet{\mathcal}
      \addtoversion{normal}{\mathcal}{cmsy}{m}{n}
    \else
    \ifnfsstwo
    \fi
  \fi
\fi

\def	\arcsec	{^{\prime\prime}}
\def	\arcsecsq	{\,{\rm arcsec}^2}	
\def	\cm	{{\rm cm}}
\def    \gtsim  {\lower.5ex\hbox{$\; \buildrel > \over \sim \;$}} 
\def    \ltsim  {\lower.5ex\hbox{$\; \buildrel < \over \sim \;$}} 
\def\H		{{\rm H}}
\def\HH		{{\rm H}_2}
\def\micron	{\mu{\rm m}}
\def\nH		{n_{\rm H}}
\def\beq	{\begin{equation}}
\def\eeq	{\end{equation}}

\def\hexnumber#1{\ifcase#1 0\or1\or2\or3\or4\or5\or6\or7\or8\or9\or
 A\or B\or C\or D\or E\or F\fi }

\makeatletter
\ifx\CUP@mtlplain@loaded\undefined
\else
\fi
\makeatother
\makeatletter
\ifx\CUP@mtlplain@loaded\undefined
\else
\fi
\makeatother
\title[Theoretical Models of PDRs]
{Theoretical Models of Photodissociation Fronts}

\author[B.T. Draine \& F. Bertoldi]
{B. T. Draine$^1$\\
\and \ns Frank Bertoldi$^2$}

\affiliation{$^1$Princeton University Observatory, 
Princeton, NJ 08544-1001, USA\\[\affilskip]
$^2$Max-Planck-Institut f\"ur Radioastronomie, D-53121 Bonn, Germany}

\begin{document}
\ifnfssone
\else
  \ifnfsstwo
  \else
    \ifoldfss
      \let\mathcal\cal
      \let\mathrm\rm
      \let\mathsf\sf
    \fi
  \fi
\fi

\maketitle

\vspace*{-15em}
\centerline{to appear in {\it H$_2$ in Space}, ed. F. Combes \& G. Pineau des
Fo\^rets (Cambridge Univ. Press), 2000}
\vspace*{14em}

\begin{abstract}
Observations of $\HH$ line emission have revealed higher-than-expected
gas temperatures in a number of photodissociation fronts.
We discuss the heating and cooling
processes in photodissociation regions.
Observations of NGC 2023 are compared to a theoretical model in which
there is substantial gas at temperatures $T=500-1000$K heated by
photoelectric emission and collisional deexcitation of $\HH$.
In general the model successfully reproduces the observed $\HH$ line
emission from a wide range of energy levels.
The observed [SiII]34.8$\micron$ emission appears to indicate substantial
depletion of Si in NGC~2023.
\end{abstract}

\firstsection 
\section{Introduction}

A significant fraction of
the ultraviolet radiation emitted by massive stars 
impinges on the molecular gas associated with star formation.
The resulting
photodissociation regions (PDRs) 
therefore play an important role in reprocessing
the energy flow in star-forming galaxies.
Modelling these PDRs is therefore an important theoretical challenge, both
to test our understanding of the physical processes in interstellar gas, and
to interpret observations of star-forming galaxies.

\begin{figure}
\centerline{\psfig{figure=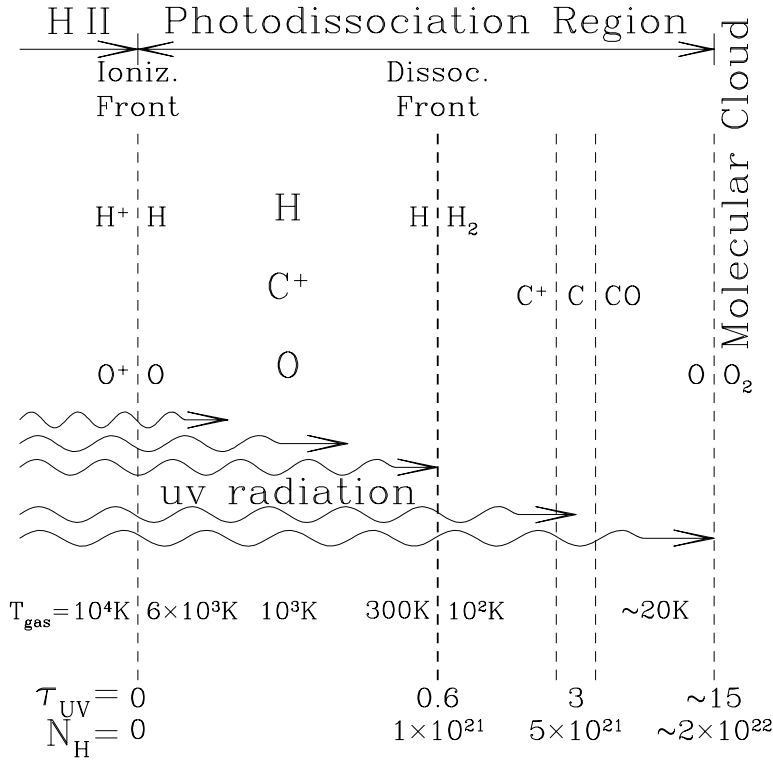,width=10.5cm,angle=0}}
\caption{Schematic diagram showing the different zones in a photodissociation
region.\label{fig:pdr_cartoon}}
\end{figure}

It is frequently the case that 
the illuminating star is hot enough to produce an H~II region,
in which case the photodissociation
region is bounded on one side by an ionization front, and on the other by
cold molecular gas which
has not yet been appreciably affected by ultraviolet radiation.
The $h\nu<13.6$eV photons propagating beyond the ionization front
raise the fractional ionization,
photoexcite and photodissociate the $\HH$,
and heat the gas via photoemission from dust
and collisional deexcitation of vibrationally-excited H$_2$.
Figure \ref{fig:pdr_cartoon} shows the different layers within a PDR.

\section{PDR Thermometry}

The lower rotational levels of $\HH$ 
tend to be in LTE (except perhaps for the
ortho/para ratio, which adjusts relatively slowly),
so we can use the relative level populations of $\HH$ as a 
PDR thermometer.

We can also
use the absolute column densities $N(v,J)$ to tell us how much molecular gas is
at different temperatures.
Because the $\HH$ quadrupole vibration-rotation lines are generally 
optically thin, only dust extinction affects the
radiative transfer.

The level populations $N(v,J)$ can be studied using vibrational transitions
(ground-based K,H,J,I,R band spectroscopy)
or $v$=0--0 pure rotation lines
(some from the ground, but 
mainly from space, with ISO or SIRTF, or from the stratosphere, with SOFIA).

\section{Modeling PDRs}

The central physical process in a PDR
is the photoexcitation
and photodissociation of $\HH$ through the Lyman and Werner band lines of
$\HH$, and it is therefore important to model this process accurately.
In our models (Draine \& Bertoldi 1996) we
explicitly include radiative transfer in the
28765 permitted Lyman and Werner band lines
between $\HH$ levels with $J\leq29$, using wavelengths and
oscillator strengths 
from Abgrall \& Roueff (1989) and Abgrall et al. (1993a,b).
The lines are treated independently assuming Voigt line profiles and
attenuation by dust, but we include a
a statistical correction for the effects of line overlap of the
$\HH$ lines
(Draine \& Bertoldi 1996).

Lyman and Werner band photons are absorbed by $\HH$ in the ground
electronic state X$^1\Sigma_g^+$, resulting in excitation to vibration-rotation
levels of either the
B$^1\Sigma_u^+$ or C$^1\Pi_u^{\pm}$ electronic states.
The electronically-excited state will decay by 
spontaneous emission of an ultraviolet
photon.
About 85\% of the downward transitions will be
to a bound $(v,J)$
level of the
ground electronic state, but about 15\% of the transitions will be
to the vibrational continuum of the ground electronic state, producing
two H atoms, typically moving apart with a kinetic energy of
a fraction of $\sim$1~eV (Stephens \& Dalgarno 1973).

Bertoldi \& Draine (1996) discussed the propagation of coupled 
ionization-dissociation fronts, and showed that except when the 
PDR is driven by radiation from a very hot star, it
is usually the case that the PDR moves into the molecular cloud slowly
enough that thermal and chemical conditions are close to being in
steady-state balance: most importantly, 
$\HH$ destruction is nearly balanced by $\HH$ formation,
and heating is nearly balanced by cooling.
It is then convenient to approximate the structure of the PDR by requiring
precise thermochemical steady state at each point.
The $\HH$ abundance is therefore determined by a balance between destruction
by photodissociation and formation on grains,
\beq
R_{gr}\nH n(\H) = \zeta_{pd} n(\HH)
\eeq
where $\zeta_{pd}$ is the local rate for photodissociation of $\HH$, and
$R_{gr}$ is the rate coefficient for $\HH$ formation on grains, with the
grain abundance assumed proportional to 
$\nH\equiv 2n(\HH)+n(\H)+n(\H^+)$.
We solve for the steady-state populations of the 299 bound $(v,J)$
levels with $J\leq29$, including UV pumping, spontaneous decay, and
collisional transitions.
The distribution over $(v,J)$ of newly-formed $\HH$ produced
on grain surfaces is uncertain; our models allow us to consider various
possibilities.

The gas is heated primarily by photoelectrons ejected from
dust grains, and (for densities $\nH\gtsim 10^4\cm^{-3}$) collisional
deexcitation of vibrationally-excited $\HH$ resulting from UV pumping.
In our models we use photoelectric heating rates estimated by
Weingartner \& Draine (2000a); we assume dust with $R_V=5.5$ with
5\% of the solar carbon in ultrasmall carbonaceous grains.

The cooling of the gas is dominated by [CII]158$\micron$,
[SiII]35$\micron$, [OI]63$\micron$, and [FeII]26$\micron$.
The [OI]63$\micron$ line is often optically thick; we use an
escape probability treatment.

\begin{figure}
\centerline{\psfig{figure=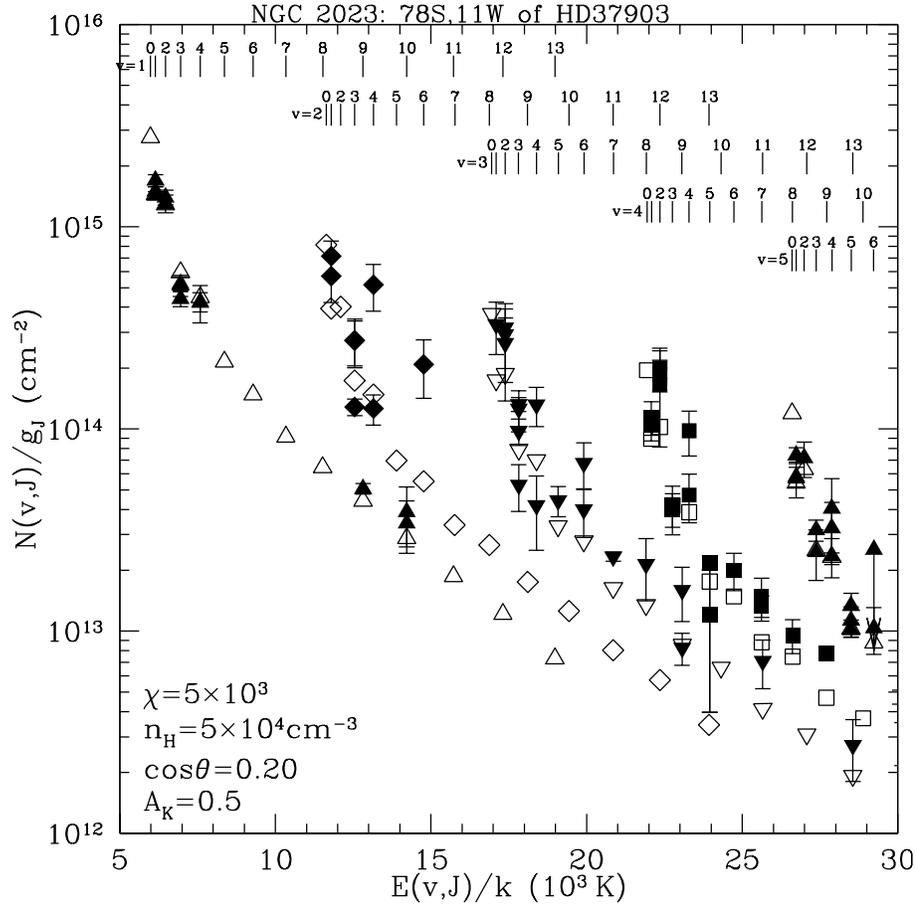,width=12cm,angle=0}}
\caption{$N(v,J)/g_J$ for vibrationally-excited levels of $\HH$ toward
	the southern bar in NGC 2023.  Open symbols: plane-parallel model
	with $\nH=5\times10^4\cm^{-3}$ and $\chi=5000$
	observed from angle $\theta=78^\circ$ rel. to surface normal.
	Filled symbols: level populations obtained from observed line
	intensities after correction for foreground extinction with
	$A_K=0.5$mag.
	Data from Burton et al.\ (1998) and McCartney et al.\ (1999).
	\label{fig:v=12345}}
\end{figure}
\begin{figure}
\centerline{\psfig{figure=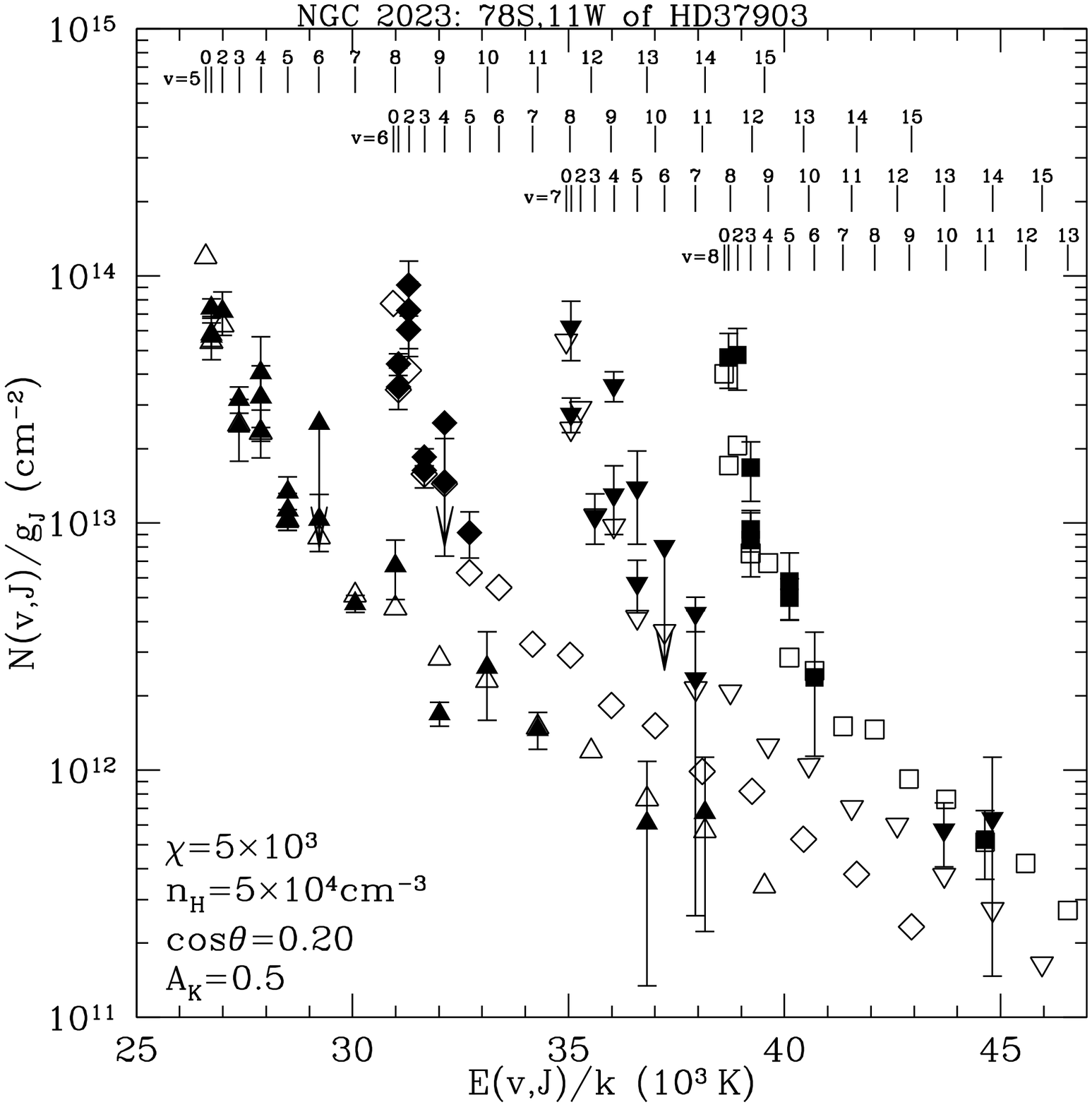,width=10cm,angle=0}}
\caption{Same as Fig.\ \ref{fig:v=12345}, but for $v=5-8$.
\label{fig:v=5678}}
\bigskip
\centerline{\psfig{figure=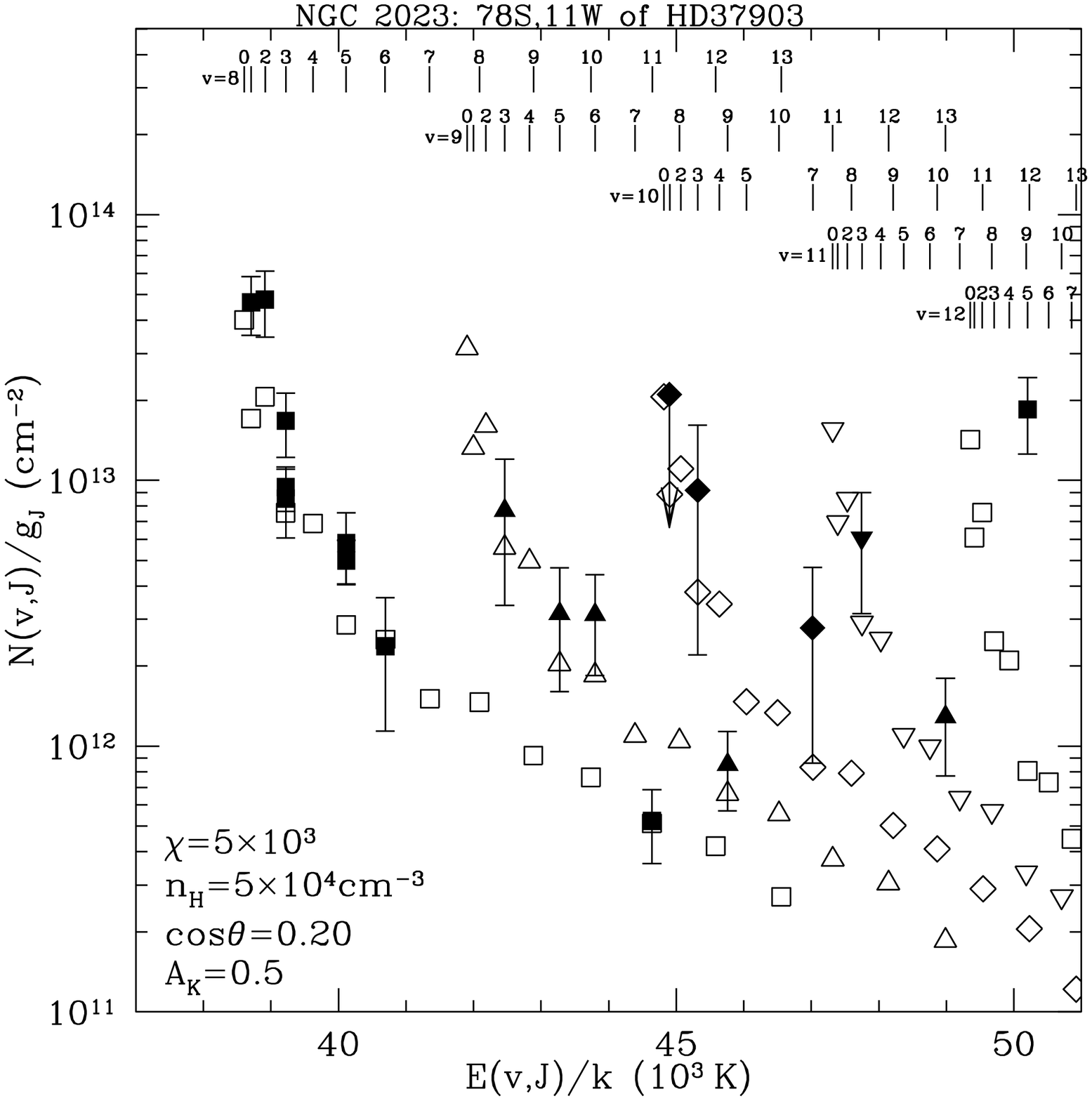,width=10cm,angle=0}}
\caption{Same as Fig.\ \ref{fig:v=12345}, 
	but for $v=8-12$.
	\label{fig:v=89101112}}
\end{figure}

\section{NGC 2023: A Test Case}

The B1.5V star HD~37903 is situated just outside the surface of the
L1630 molecular cloud, resulting in the famous reflection nebula NGC 2023.
Many emission lines of $\HH$ have now been measured for NGC 2023,
making it an ideal test for PDR models.
High resolution images of NGC 2023 in the 1-0S(1) line have recently
been published (Rouan et al.\ 1997; Field et al.\ 1998;
McCartney et al.\ 1999), showing pronounced filamentary structure, notably
strong emission along the ``southern bar'' 78$\arcsec$ south of HD~37903.

Black \& van Dishoeck (1987) produced the first models to explain the
$\HH$ fluorescence from NGC 2023.  Their favored 
model assumed cold $\HH$ with a
density $\nH\approx10^4\cm^{-3}$ and
an ultraviolet radiation field enhanced by a factor
$\chi\approx700$ relative to the Habing (1968) intensity.
Draine \& Bertoldi (1996) argued that some warm gas with $T\approx 500 - 1000$K
with density $\nH\approx10^5\cm^{-3}$ 
was required to explain the observed line ratios.

New observations have become available both from the ground
(Burton et al.\ 1998; McCartney et al.\ 1999) and from ISO
(Bertoldi et al.\ 2000a).
From ground-based spectrophotometry of the bright ``southern emission bar''
it is now clear that 
the foreground extinction at K is
$A_K\approx0.5$mag, implying a far-red extinction
$A_{0.8\micron}\approx 2.7$mag.
For each observed $\HH$ emission line we compute the
column density $N(v,J)$ in the excited state 
required to reproduce the observed surface brightness,
for an assumed foreground extinction $A_K=0.5$mag.

A detailed discussion of our modelling of NGC 2023 
will be reported elsewhere, but here
we show one example of a model developed to reproduce
the observations of the southern emission bar.
We assume the incident radiation field to be
enhanced by a factor $\chi\approx5000$ relative to the Habing (1968) flux,
and a gas density $\nH=5\times10^4\cm^{-3}$.
The PDR is assumed to be plane-parallel, inclined relative to the
line of sight with $\cos\theta=0.2$, where $\theta$ is the angle between
the normal to the PDR (= the direction from the PDR to the illuminating star)
and the direction to the Earth.

In Figures \ref{fig:v=12345}--\ref{fig:v=89101112} 
we show $N(v,J)/g_J$ for the model together with
the observed level populations, versus the energy $E(v,J)$ of the
excited state.
Open symbols are model results; filled symbols are observations.
The model is seen to be in quite good agreement with the populations of
the vibrationally-excited levels all the way up to $v=12$!
The generally excellent agreement between model and observations indicates that
the description of the ultraviolet pumping process is basically sound.
The fact that for a given vibrational level the rotational level populations
appear to be in good agreement with observations indicates that the
model has approximately the correct temperature and density.
Only for a few weak lines are the reported fluxes very different from the
model predictions: 12-6~Q(5)$\,\lambda=0.8225\micron$, where the
reported flux is a factor of 20 stronger than expected,
and 9-4~S(11)$\,\lambda=0.7663\micron$, where the observed flux is
a factor $\sim$7 stronger than expected.
With so many other lines in good agreement, perhaps these lines have been
misidentified.  Note, however, that these are the only two lines
from levels with $E(v,J)/k>48\times10^3$K, so perhaps there are processes
populating these very high energy levels which have not been included in
the model (see Bertoldi et al.\ 2000b).

\begin{figure}
\centerline{\psfig{figure=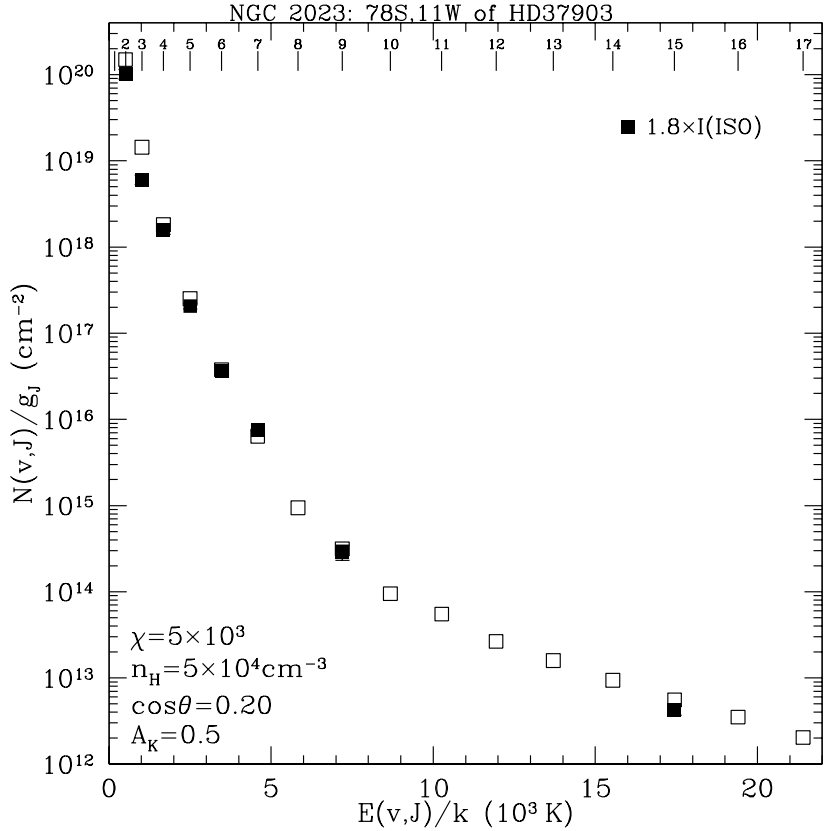,width=11.cm,angle=0}}
\caption{Level populations for rotationally-excited levels of the
	ground vibrational state in NGC 2023.  Beam-averaged column densities
	observed by ISO have been multiplied by 1.8 to allow for likely
	enhancement of the surface brightness on the southern bar.
	\label{fig:v=0}}
\end{figure}

\begin{figure}
\centerline{\psfig{figure=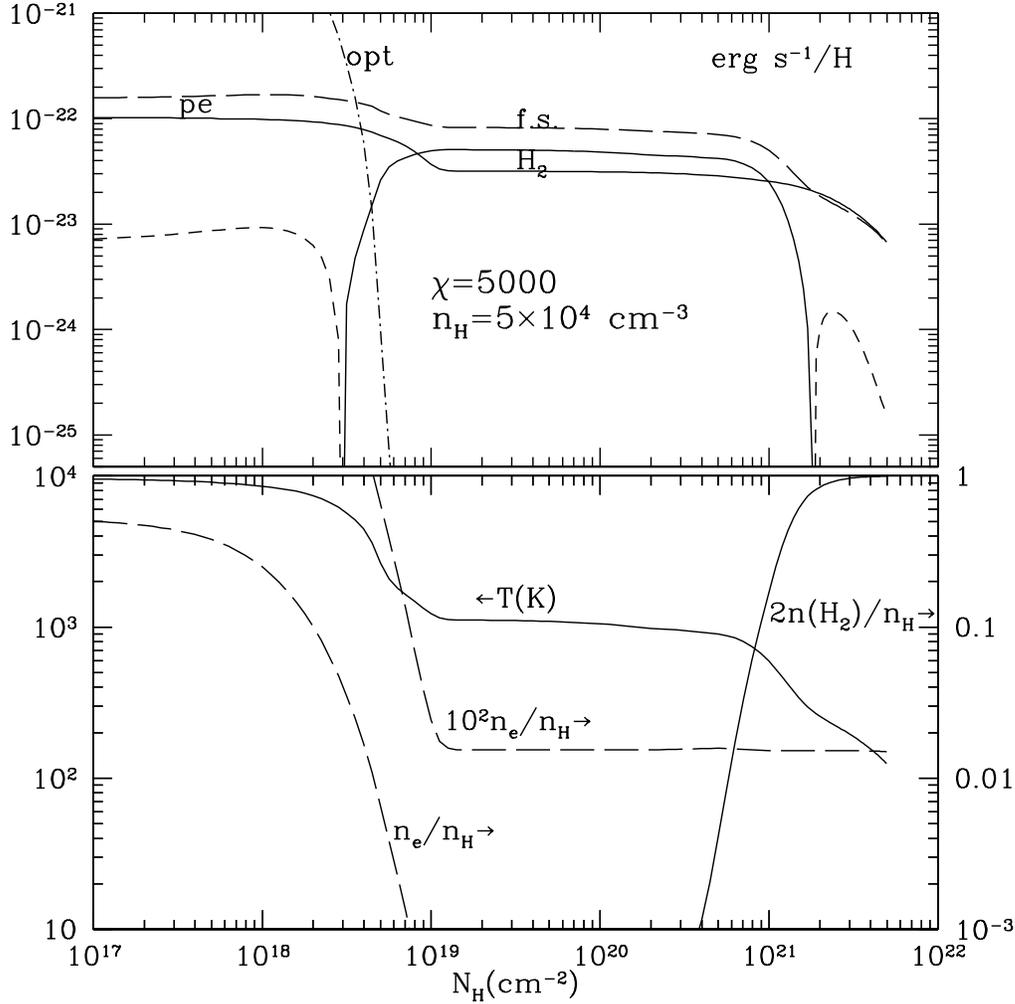,width=13.5cm,angle=0}}
\caption{Lower panel: temperature $T$, ionization fraction
	$n_e/\nH$, and $\HH$
	fraction as function of column density $N_\H$ measured from the
	ionization front.
	Upper panel: Heating or
	cooling contributions due to photoelectric emission from dust (p.e.),
	fine structure emission (f.s.),
	optical line emission (opt),
	and collisional excitation/deexcitation of $\HH$ ($\HH$).
	Solid lines indicate heating; broken lines indicate cooling.
	$\HH$ heats the gas over the region
	$5000\gtsim T\gtsim 300$K.
	\label{fig:2023model}}
\end{figure}

Figure \ref{fig:v=0} shows the $v=0$ level populations for our model, together
with the level populations inferred from the ISO observations
toward NGC 2023 (Bertoldi et al.\ 2000a), except that
the surface brightnesses measured by the relatively large ISO beam have
been arbitrarily increased by a factor 1.8 -- the reasoning here is that
the bright southern bar represents a region with unusually high 
limb-brightening, probably about a factor of 2 higher than the average
over the ISO beam.
With this adjustment, we obtain quite good agreement for most of the
rotational levels; the largest discrepancy is for $J-3$, where the model
surface brightness is about twice the observed value.
Note that the observed emission from the $J=15$ level is in excellent
agreement with the model!
The level populations obviously depart strongly from a single-temperature
fit; this is mainly due to the range of gas temperatures present
(see Figure \ref{fig:2023model}) and the excitation of newly-formed
$\HH$ leaving grain surfaces.


The model is shown in Figure \ref{fig:2023model}.
Photoelectric emission from dust grains dominates the heating over much of
the PDR, but
for $10^{19}\ltsim N_\H \ltsim 10^{21}\cm^{-2}$
the dominant heat source is collisional deexcitation
of $\HH$ which has been vibrationally excited by UV pumping.
The combination of photoelectric heating and heating by deexcitation of
$\HH$ manages to maintain a gas temperature of $\sim10^3$K over a substantial
zone, with the gas temperature still as high as $\sim700$K when the
$\HH$ fraction has risen to 10\%.

The model assumes gas phase abundances C/H=$1.4\times10^{-4}$
(1/3 solar),
O/H= $3.2\times10^{-4}$ (1/2 solar),
Si/H=$9.0\times10^{-7}$ (1/40 solar),
Fe/H=$1.3\times10^{-7}$ (1/250 solar).
Predicted fine structure line intensities for the bright southern bar
(for limb-brightening $1/\cos\theta=5$) 
are given in Table \ref{tab:fs}.
Most of the lines are not yet observed, but the intensity measured by ISO
for [SiII]34.8$\micron$ is only 1/25 of the model prediction!  Part of
the discrepancy can be attributed to beam dilution: the ISO beam is
660$\arcsecsq$, whereas the bright bar is only $\sim$50$\arcsecsq$.
However, there presumably is [Si~II] emission away from the bar, so it
appears likely that the Si is depleted by $\sim$400 relative
to solar.
It is interesting to compare this with the S140 PDR, where the
inferred Si abundance was 1/40 solar
(Timmermann et al.\ 1996).
Walmsley et al.\ (1999) have recently discussed the puzzling variations in
Si abundance observed in PDRs.

\section{Discussion}

There are a number of areas for improvement in our modelling of PDRs:
\begin{itemize}
\item We need accurate cross sections
for collisional excitation and dexcitation of $\HH$.
\item We need to understand the details of the $\HH$ formation process
in PDRs -- the overall rate and the $(v,J)$ distribution of newly-formed $\HH$.
\item The rate of photoelectric heating by dust is critical, particularly
since drift of the dust grains can alter the dust/gas ratio in the PDR
(Weingartner \& Draine 1999, 2000b).
\item Finally, we need more accurate spectroscophotometry, with refined
angular resolution, to allow us to understand the geometry of the filaments
and sheets which are apparent in high-resolution images of NGC 2023 and other
PDRs.
\end{itemize}

\begin{acknowledgments}
This research was supported in part by NSF grant AST96-19429.
\end{acknowledgments}

\begin{table}
\begin{center}
\begin{tabular}{lccl}
line	& predicted & observed\\
$[{\rm SiII}]34.81\micron$	&$1.5\times10^{-3}$	&$5.7\times10^{-5}$
	&erg~cm$^{-2}$~s$^{-1}$~sr$^{-1}$\\
$[{\rm FeII}]25.99\micron$	&$1.8\times10^{-4}$	&-\\
$[{\rm OI}]63.18\micron$	&$4.5\times10^{-2}$	&-\\
$[{\rm OI}]145.5\micron$	&$2.9\times10^{-3}$	&-\\
$[{\rm CII}]157.7\micron$	&$4.6\times10^{-3}$	&-\\
\end{tabular}
\caption{NGC 2023 Fine Structure Line Emission\label{tab:fs}}
\end{center}
\end{table}

\end{document}